\documentstyle[12pt]{article}
\input epsf.tex

\renewcommand{\theequation}{\thesection.\arabic{equation}}
\newcommand{\be}{\begin{equation}}   \newcommand{\ee}{\end{equation}}
\newcommand{\bear}{\begin{eqnarray}}
\newcommand{\eear}{\end{eqnarray}}
\newcommand{\ba}{\begin{array}}      \newcommand{\ea}{\end{array}}
\newcommand{\lae}{\begin{array}{c}\,\sim\vspace{-21pt}\\< \end{array}}
\newcommand{\gae}{\begin{array}{c}\,\sim\vspace{-21pt}\\> \end{array}}

\def\vbr{$\vphantom{\sqrt{F_e^i}}$}

\begin{document}

\pagestyle{empty}
\begin{titlepage}
\def\thepage {}        

\title{\Large \bf
Gauge Coupling Unification with Extra Dimensions
\\ [2mm]
and Gravitational Scale Effects \\ [1cm]}

\author{\bf \large  Hsin-Chia Cheng$^1$, Bogdan A.~Dobrescu$^1$, and
Christopher T.~Hill$^{1,2}$ \\
\\
{\small {\it $^1$Fermi National Accelerator Laboratory}}\\
{\small {\it P.O. Box 500, Batavia, Illinois, 60510, USA \thanks{e-mail
  addresses: hcheng@fnal.gov, bdob@fnal.gov, hill@fnal.gov} }}\\
\\
{\small {\it $^2$ The University of Chicago }}\\
{\small {\it Chicago, Illinois, 60510, USA }}\\ }

\date{ }

\maketitle

   \vspace*{-11.4cm}
\noindent
\makebox[10.7cm][l]{hep-ph/9906327} FERMILAB-PUB-99/168-T \\ [1mm]
\makebox[10.7cm][l]{June 10, 1999} 
\\

 \vspace*{11cm}

\baselineskip=18pt

\begin{abstract}

   {\normalsize
We study gauge coupling unification in the presence of
extra dimensions compactified at a few TeV.
Achieving unification requires a large number of
gauge boson Kaluza-Klein excitations lighter
than the string scale, such that the
higher-dimensional gauge couplings are ${\cal O}(1)$
in string scale units.
Corrections to the gauge couplings from two or more loops
are about 10\% or larger, hence string (or M) theory is
generally expected to be strongly coupled in $\sim$TeV-scale
extra-dimensional scenarios.
Higher-dimensional operators induced by quantum gravitational effects 
can shift the gauge couplings by a few percent.
These effects are sufficiently large that even the minimal
Standard Model, or the MSSM, allow unification at a scale in 
the $\sim$TeV range. The strongly coupled unified theory may induce 
dynamical electroweak symmetry breaking.
}

\end{abstract}

\vfill
\end{titlepage}

\baselineskip=18pt
\pagestyle{plain}
\setcounter{page}{1}

\section{Introduction}

A striking feature of the observed elementary fermions is that
they fit into complete $SU(5)$ representations \cite{geogla}.
It is therefore compelling to try to promote this $SU(5)$
to a gauge symmetry which is broken down to the Standard Model
gauge group. In the usual scenario, a necessary condition is
that the $SU(3)_C \times SU(2)_W \times U(1)_Y$ gauge couplings
unify at one energy scale. The measured gauge couplings seem to be
roughly converging if their logarithmic running is extrapolated over
fourteen orders of magnitude in energy scale. This running is,
of course,  sensitive to the full elementary
field content above the $M_Z$ scale, so that gauge coupling
unification is a model dependent issue.

In the Minimal Supersymmetric Standard Model (MSSM) where superpartners
have masses close to the electroweak scale, the
gauge couplings converge at a grand unification (GUT) scale of
$M_{\rm GUT} \sim 2 \times 10^{16}$ GeV
with a precision of a few percent~\cite{mssm,precision}.
Quantitatively, given the precisely measured $SU(2)_W$ and $U(1)_Y$
couplings, requiring that the 3 gauge couplings meet exactly at one
point (``naive" unification)
gives a prediction of $\alpha_3(M_Z) \approx 0.129$, with a mild dependence
on the superpartner masses.
This result is higher than the experimental average
\cite{alphas} $\alpha_3(M_Z) \approx 0.119 \pm 0.002$ by about 4 or 5 standard
deviations.\footnote{The most detailed current analyses of perturbative GUT
unification in the MSSM are given in \cite{precision}; at the time of the
publication of these papers there was some conflict between the
values for $\alpha_3(M_Z)$ extracted from $Z$ pole observables and from
low energy data, respectively.
Currently, these two values are in better agreement, so that
the error bars on the world average of $\alpha_3(M_Z)$ have been reduced,
which widens the discrepancy with the predictions of naive unification. }
However, it is natural to expect some  corrections of order
a few percent at the GUT scale
due to threshold effects~\cite{threshold,precision}. Allowing for
such uncertainties  the 3 gauge couplings of the MSSM can unify.

Are other theories disfavored if they do not unify with the same precision
as the MSSM?  Unfortunately, we cannot answer this question yet.
Gauge coupling unification offers
no observed phenomenon that measures the coupling constants
at $M_{\rm GUT}$ ({\it e.g.,} if proton decay is observed
we will have in principle a relationship of the effective
coupling constant at $M_{\rm GUT}$ to the low energy couplings
and a definitive observation of the phenomenon of unification).
We do not know that the couplings  should meet identically at $M_{\rm GUT}$
or whether there may be other larger, decoupled
effects which renormalize the couplings
in unknown ways, which may imply drastically different scenarios at
$M_{\rm GUT}$ than, e.g., the MSSM \cite{Hill84}.

Recently, Dienes, Dudas and Ghergheta \cite{ddg} have considered lowering the
unification scale in the MSSM by allowing the gauge fields to propagate
in extra dimensions which are presumed to open up at low
energy scales, such as the $\sim$TeV scale \cite{Antoniadis:1990ew}.
Above the compactification scale of these extra
dimensions, the gauge couplings depend on the energy scale as a
power law, so that they converge very quickly.
It is interesting that the gauge couplings still unify approximately in
the presence of the extra dimensions~\cite{ddg,uni,Perez-Lorenzana:1999qb}.
It may even be possible that the unification scale is not far above
the electroweak scale. In that case, the unification of the gauge couplings
may have a chance to be probed directly in future experiments.

An immediate problem of a low energy unification scale is
proton stability. In string theory, there can be gauge coupling
unification even without the $SU(5)$ gauge symmetry, but the
higher excitations of the $X$ and $Y$ gauge bosons may still be present
and hence mediate proton decays with an unacceptable rate.
Nevertheless, there are potentially higher dimensional
theoretical solutions
to this problem~\cite{ddg,AS}. For example, in string theory,
the light fermions do not necessarily come from the same generations
at the string scale. Moreover, if the quarks and leptons
are localized at different
points (on different branes) in the extra dimensions,
then the proton decay rate can be highly suppressed.

The low-scale extra-dimensional theories are at present
somewhat theoretically unconstrained.  This is due to the fact
that a large number of gauge invariant irrelevant (higher mass-dimension)
operators can arise at the $\sim$ TeV scale and lead to a
variety of effects, some of which obviate the constraints on
the Standard Model. For example,
Hall and Kolda \cite{HK} (see also \cite{hmass})
have argued that the $S$-$T$ constraints
of the oblique electroweak radiative corrections, which favor
a low mass Higgs boson (and no additional chiral multiplets),
disappear in the presence of certain allowed higher dimensional
operators.

The present paper studies the corrections to the gauge couplings induced
by the  Kaluza-Klein (KK) excitations and by the higher-dimension operators
which can occur in the presence of extra dimensions and a low string 
scale.\footnote{Throughout this paper, by extra dimensions we mean compact
dimensions accessible to the gauge bosons. A low string
scale, which coincides with the unification scale, may require
larger dimensions accessible only to the gravitons.}

A key point of this paper is that, in order to achieve unification
and bring the three gauge couplings together
at a scale in the TeV range, there is need for many KK
excitations contributing to the running. Therefore, 
the effective coupling constants, or equivalently the higher-dimensional
gauge couplings above the compactification scale, are {\em  necessarily large}.
It is in fact remarkable that
at the scale where the three gauge couplings appear to converge at
the one-loop order,
the loop expansion parameter is usually of order one, or slightly smaller.
This indicates that the underlying (string or M) theory is in the
non-perturbative regime, which is in agreement with a general argument
that the string coupling is likely to be of order one \cite{dine}.
In this case one does not know how to compute the string scale corrections
to the Standard Model gauge couplings, and therefore the gauge unification
is more uncertain.
We adopt an effective field theory below $M_s$,
in which there are string scale suppressed operators with coefficients
of order one (or smaller if there are approximate global symmetries)
determined by the non-perturbative string dynamics.

This can dramatically alter our
intuition about physics beyond
the low energy Standard Model.
For example, in the minimal (non-supersymmetric) Standard Model it has
long been argued that (1)
the gauge couplings do not unify because they fail to meet at
a common value by a large margin, and (2) because of the hierarchy
problem, it is unlikely that we can extrapolate the gauge couplings
to very high energy scales without introducing new physics at the
TeV scale anyway.   However,
the strong dynamics at the low energy
string scale in these novel scenarios
offers the interesting possibility that the fundamental scale,
which we take to be the string scale,
$M_s$, is close to the electroweak scale~\cite{lykken,largedim,warp},
so that there is no hierarchy. In the end,
these scenarios predict a new strong dynamics at the $\sim$TeV scale.
It is therefore worthwhile to reexamine the
general question of unification in
the context of extra dimensions at the TeV
scale and strong dynamics.

Our paper is organized as follows.
In Section 2 we discuss in general the scale dependence of the
$SU(3)_C \times SU(2)_W \times U(1)_Y$ gauge couplings in the presence
of extra dimensions accessible to the gauge bosons, with a compactification
scale below $M_s$.
In Section 3 we show that any scalar field with a vacuum expectation value
(VEV) modifies the gauge couplings due to certain dimension-six
operators (assumed to be induced with a coefficient of order one times
$M_s^{-2}$ by the string dynamics).
In the Standard Model, the tree level
shifts in the gauge couplings due to the Higgs VEV are of order
$\sim \langle H \rangle^{2}/M_s^{2}$.
A large shift, linear in $\langle \phi \rangle$, occurs
when the $\phi$ scalar is an adjoint under a
unified gauge group \cite{Hill84},
or a gauge singlet.
We also show that in a non-supersymmetric theory, any scalar field
produces at one-loop
a shift in the gauge couplings of order a few percent.
Generically, the corrections to the three Standard Model
gauge couplings are different, so that models in which
the gauge couplings would not unify naively
could in fact lead to unification in the
presence of the higher-dimensional operators.

In Section 4 we present some simple examples.
Since it is not known whether the non-perturbative string dynamics
preserves some of the supercharges at $M_s$, we discuss gauge coupling
unification in models with supersymmetry broken either at $M_s$ or at
the electroweak scale.
In Section 4.1 we show that the Standard Model with extra
dimensions compactified at some TeV scale is consistent with
gauge coupling unification at the string scale provided the inverse coupling
constants receive corrections of order 5\% beyond the one-loop running.
Shifts of this size are naturally
given by dimension-six operators involving the Higgs field and the
gauge field strengths. Also
we estimate the corrections to the gauge couplings from two or more loops
to be of order 10\% in the case of one extra dimension, and larger
for more dimensions.

Since the loop expansion breaks down for more extra dimensions,
it is interesting to study what non-perturbative phenomena may occur
in this case. A possible effect
of the non-perturbative phenomena at $M_s$ is chiral symmetry breaking
in the quark sector. This leads to the existence of a composite
Higgs sector \cite{composite}, in which the scalars are
made up of quarks bound together by KK
excitations of the gluons. In Section 4.2 we point out that the ideas
of gauge coupling unification and Higgs compositeness are perfectly
compatible in the presence of extra dimensions.

In Section 4.3 we turn to supersymmetric models in extra dimensions.
Supersymmetric operators induced at the string scale give
corrections to the gauge couplings of order the $SU(3)_C \times
SU(2)_W \times U(1)_Y$ invariant combinations of VEVs suppressed
by the appropriate power of $M_s$. In the MSSM these corrections are
likely to be below 1\%, while in the Next-to-Minimal Supersymmetric
Standard Model (NMSSM) they may be as large as 10\%.
Although the one-loop running gives a somewhat better convergence of the
gauge couplings in the MSSM than in the Standard Model, the 
expansion parameters are larger and hence the perturbative series are less 
reliable in the MSSM. In fact, if
the gauge couplings unify in the MSSM with more than one compact dimensions,
then this happens in the strong coupling regime.

Section 5 includes our conclusions.
In the Appendix we comment on the current bounds on the compactification
and string scales from collider experiments and precision low energy data.

\section{Gauge Coupling Running in Extra Dimensions}
\setcounter{equation}{0}

We begin with a general discussion of the running of the
$SU(3)_C \times SU(2)_W \times U(1)_Y$ gauge couplings
assuming that there are $\delta$ compact spatial dimensions of radius
$R \ge {\cal O}($TeV$^{-1})$ which are accessible to the
gauge bosons.

The higher-dimensional field theory is non-renormalizable.  This
is a red-herring because point-like
quantum field theory is no
longer a good description for physics above the string scale, $M_s$.
The effects of the compact dimensions
can be described, nonetheless, in the four-dimensional
Minkowski spacetime by introducing  a tower of
KK modes with masses between $1/R$ and $M_s$.

The relative normalization of the three coupling constants, $\alpha_i$,
is, as usual, model dependent. In the $SU(5)$ GUT, $\alpha_3 = g_s^2/(4\pi)$,
$\alpha_2 = g^2/(4\pi)$, and $\alpha_1 = (5/3) g^{\prime 2}/(4\pi)$,
where $g_s$, $g$ and $g^\prime$ are the usual Standard Model gauge couplings.
Of course, it is not known whether the $SU(5)$ normalization is
the correct one.
Different normalizations may be imposed at the string level. For example,
non-trivial compactifications give rise to different Kac-Moody levels for the
three gauge groups~\cite{kacmoody}.
Also, the $\alpha_i$'s are normalized differently if
the $SU(3)_C$, $SU(2)_W$ and $U(1)_Y$ groups are associated with different
branes \cite{difbrane}.
Clearly, in order to decide that gauge coupling unification does not
occur in a particular model, one would have to argue that the normalizations
that lead to unification are unlikely to be given by string theory.
Making such an argument does not appear to be feasible, at least for now,
given that string or M theory may be in the strong coupling regime
\cite{dine}.

In this paper we use only the usual $SU(5)$ normalization, because this is
the most natural choice.
The experimental values of the inverse coupling constants in the
$\overline{MS}$ scheme are \cite{alphas}
\bear
\alpha_1^{-1}(M_Z) & = & 58.98 \pm 0.04
\nonumber \\ [2mm]
\alpha_2^{-1}(M_Z) & = & 29.57 \pm 0.03
\nonumber \\ [2mm]
\alpha_3^{-1}(M_Z) & = & 8.40 \pm 0.14
\eear

The $\overline{MS}$ gauge coupling constants at a scale $\mu > 1/R$ are
related to the measured coupling constants at the $Z$ pole by
\be
\alpha_i^{-1}(\mu) = \alpha_i^{-1}(M_Z)
- \frac{b_i}{2\pi} \ln \left(\frac{\mu}{M_Z}\right)
- \frac{\tilde{b}_i}{2\pi} {\cal F}(\delta, R \mu) + \Delta_i^{\rm loops}~.
\label{run}
\ee
The $b_i$ ($i = 1,2,3$) are the one-loop $\beta$-function coefficients
of the four-dimensional zero-modes
(they incorporate the threshold corrections due to
particles heavier than the $Z$, such as the top quark),
while the $\tilde{b}_i$ correspond to one KK
excitation for each field propagating in extra dimensions.
The function ${\cal F}(\delta, R \mu)$ sums the one-loop
contributions from all
the KK excitations, and $\Delta_i^{\rm loops}$ are the corrections from
two and more loops.

Let us label the KK levels by $n \ge 1$, their
masses by $M_n$ (with $M_n < M_{n+1}$), and their degeneracies by $D_n$.
In the $\overline{MS}$
scheme\footnote{In ref.~\cite{ddg} the wave function renormalization
is computed with an explicit cut-off such that in addition to the leading
logarithmic divergent terms given in eq.~(\ref{calf1}), some finite
corrections are also included. For the experimental values
$\alpha_i(M_Z)$ which correspond to the $\overline{MS}$ scheme,
the procedure used in \cite{ddg} introduces some
errors compared to eq.~(\ref{calf1}). However, for $(R \mu)^\delta \gg 1$,
these two procedures give approximately the same result, {\it i.e.}
a power law running \cite{ddg, Perez-Lorenzana:1999qb}:
${\cal F}(\delta, R \mu) \approx
2 \pi^{\delta/2}[\Gamma(\delta/2) \delta^2]^{-1}
(R \mu)^\delta$.},
\be
{\cal F}(\delta, R \mu) = \sum_{n=1}^{n(\mu)} D_n
\ln\left(\frac{\mu}{M_n}\right) ~,
\label{calf1}
\ee
where $n(\mu)$ is a defined by  $M_{n(\mu)} < M_{n(\mu)+1}$.

The total number of KK levels below the string scale, $n_{\rm max}$,
is fixed by the value of $R M_s$.
The KK mass levels $M_n < M_s$ are determined by the condition that
the equation
\be
\sum_{j = 1}^{\delta} k_j^2 = \left( M_n R \right)^2 ~,
\ee
where $k_j$ are integer variables, has at least one solution.
The degeneracy $D_n$ is the number of solutions to this equation.
The total number of KK modes is given by
\be
N_{\rm KK} = \sum_{n=1}^{n_{\rm max}} D_n ~.
\ee

For the case of only one compact dimension of radius $R > M_s^{-1}$
accessible to the gauge bosons, $n_{\rm max}$ is the
integer satisfying $n_{\rm max} \leq R M_s < n_{\rm max} + 1$, the
number of KK modes is $N_{\rm KK} = 2 n_{\rm max}$,
and $D_n = 2$. It follows that\footnote{This agrees with eq.~(B.1)
in ref.~\cite{ddg} except for an erroneous $\ln \left(M_s/M_Z\right)$ term.}
\be
{\cal F}(\delta=1, R M_s) = N_{\rm KK} \ln\left(R M_s\right)
- 2 \ln\left(n_{\rm max}!\right) ~.
\label{fdel1}
\ee

For $\delta \geq 2$, the KK levels are no longer equally spaced
and their degeneracies are level-dependent. Therefore,
${\cal F}(\delta \geq 2, R M_s)$ is given by an expression similar with
(\ref{fdel1}), but with the second term on the right-hand-side modified
due to the non-uniform KK levels.
In Table 1 we list the KK mass levels and degeneracies
that saturate the bound ${\cal F}(\delta \geq 2, R M_s) \lae 42$ (this
value is relevant for the MSSM, see Section 4.3).

\begin{table}[ht!]
\centering
\renewcommand{\arraystretch}{1.5}\small
\begin{tabular}{@{}l||c||c|c|c|c| c| c| c| c| c| c|c| c| c|c|@{} } \cline{2-16}
 &  $ n$ & $\; 1\; $ & 2 & 3 & 4 & 5 & 6 & 7 & 8 & $ 9 $ & 10
& 11& 12& 13& 14
\\ \hline \hline\vbr\vbr
 \vline \hfill $\; \delta = 2 \!\!$  &
$M_n R$ & 1 & $\sqrt{2}$ & 2 & $\sqrt{5}$ & $\sqrt{8}$ & 3 &
 $\!\sqrt{10}\!$ &      $\! \sqrt{13}\!$
& 4 & $\!\sqrt{17}\! $ & $\! \sqrt{18}\! $ & $\!\sqrt{20}\!$ & $5$
& $\!\sqrt{26}\!$ \\ \cline{2-16}
\vline \hfill  & $D_n$  & 4 & 4 & 4 & 8 & 4 & 4 & 8 & 8 & 4 & 8 & 4
& 8 & 12 & 8
\\ \hline
\vline \hfill  $\; \delta = 3 \!\!$ &
$M_n R$ & 1 & $\sqrt{2}$ & $\sqrt{3}$ & 2 & $\sqrt{5}$ & $\sqrt{6}$
& $\!\sqrt{8}\!$ & 3 &  $\!\sqrt{10}\!$
\\ \cline{2-11}
\vline \hfill   & $D_n$  & 6 & 12 & 8 & 6 & 24 & 24 & 12 & 30 & 24
\\ \cline{1-11}
\vline \hfill  $\;  \delta = 4 \!\!$ &
$M_n R$ & 1 & $\sqrt{2}$ & $\sqrt{3}$ & 2 & $\sqrt{5}$ & $\sqrt{6}$
\\ \cline{2-8}
\vline \hfill   & $D_n$  & 8 & 24 & 32 & 24 & 48 & 96
\\ \cline{1-8}
\vline \hfill  $\; \delta = 5 \!\!$ &
$M_n R$ & 1 & $\sqrt{2}$ & $\sqrt{3}$ & 2
\\ \cline{2-6}
\vline \hfill   & $D_n$  & 10 & 40 & 80 & 90
\\ \cline{1-6}
\vline \hfill  $\;  \delta = 6 \!\!$ &
$M_n R$ & 1 & $\sqrt{2}$ & $\sqrt{3}$
\\ \cline{2-5}
\vline \hfill   & $D_n$  & 12 & 60 & 160
\\ \cline{1-5}
\end{tabular} 
\parbox{15.cm}{\small \caption{\small
The masses, $M_n$, and degeneracies, $D_n$, of the KK levels
for $2 \leq \delta \leq 6$ compact dimensions. For the levels shown here,
${\cal F}(\delta, R M_n) < 42$.
\label{KKlevels} } }
\end{table}

So far we have discussed only the one-loop running of the gauge couplings.
In addressing the question of gauge coupling unification one has
to decide whether the perturbative expansion is convergent, and, if it is,
to find how large are the higher-loop corrections
$\Delta_i^{\rm loops}$. This is especially important
in the presence of a large number of KK modes, because the
effective gauge coupling is given by the 't Hooft coupling $N_{\rm KK} g^2$.
More precisely, the loop-expansion parameter for the $SU(3)_C$ gauge group
at a scale $R^{-1} < \mu < M_s$ is given by
\be
\sim N_c N_{\rm KK}(\mu) \frac{\alpha_3(\mu)}{4\pi} ~,
\ee
where $N_c = 3$ is the number of colors, and the $4 \pi$ suppression
is due to the integration over the angular variables.
Of course, the numerical factor that multiplies this expansion parameter
is not known, so that we cannot decide exactly whether the loop series
is convergent when the expansion parameter is not significantly smaller
than one.
Since the size of the higher-loop corrections is model dependent,
we will discuss them (in Section 4.1)
in a simple example: the Standard Model in extra dimensions.

\section{Effects of Quantum Gravity on Gauge Couplings}
\setcounter{equation}{0}

The gauge coupling in the $(4+\delta)$-dimensional
theory and the effective 4-dimensional gauge coupling are related by a
volume ratio factor (or equivalently, the number of the KK states,
$N_{\rm KK}$)~\cite{ADD2}.
More precisely, the Standard Model squared gauge couplings
in $(4+\delta)$-dimensions are given by $N_{\rm KK} g_i^2/M_s^{\delta}$.
A large number of KK modes corresponds to a large $(4+\delta)$-dimensional
coupling, indicating that the theory may become non-perturbative at
the string scale.
Other arguments \cite{dine}, based on entirely different reasons,
also suggest that the string coupling is of order one.

Hence, the string scale corrections to the gauge couplings could
be large. The usual perturbative computations of these corrections
in string theory \cite{kacmoody, moreuni} may not be trusted here due to the
large string coupling. We will use an effective field theory at $M_s$
in which we parameterize the string effects by higher-dimensional
gauge invariant operators with arbitrary coefficients.
Generically, we expect these coefficients to be of order one
(or smaller if some global symmetries are approximately preserved)
times the appropriate power of $M_s$.

In this section we study the possible corrections to the gauge couplings
due to higher-dimensional operators.

\subsection{Shifts in gauge couplings due to vacuum expectation values}

If the effective field theory below the scale $M_s$
includes a scalar field, $\phi$,
then the following operator in the (effective) 4-dimensional theory
is induced in the Lagrangian at $M_s$:
\be
\sum_{i = 1}^{3} \frac{C_i}{M_s^{2}}
\phi^\dagger \phi F^{\mu\nu}_i F_{i\,\mu\nu} ~,
\label{phiphi}
\ee
where $F^{\mu\nu}_i$, $i = 1,2,3$  are the
$U(1)_Y$, $SU(2)_W$ and $SU(3)_C$ gauge field strengths, and $C_i$
are real dimensionless coefficients determined by the string dynamics.
$C_i$ are expected to be of order
one\footnote{One may wonder whether $C_i$ should contain the gauge couplings
squared, $g_i^2$, as it would appear after rescaling the gauge
kinetic terms to the canonical form if their normalization
was initially $- 1/(4 g_i^2) F^{\mu\nu}_i F_{i \, \mu\nu}$.
In practice, there is little difference between the
two normalizations because the gauge coupling is expected to be of order one
at $M_s$, and this uncertainty is included in the
statement that $C_i \sim {\cal O}(1)$.}
 at the scale $M_s$, if
$\phi$ does not propagate in the extra dimensions
at scales below $M_s$ [otherwise $C_i$ are suppressed by $(R M_s)^\delta$].
If the GUT symmetry is broken at $M_s$, then there is no
reason to expect the three $C_i$ to be equal.
In the Standard Model, $\phi$ can be the Higgs doublet.

If $\phi$ has a non-zero VEV, then the operator (\ref{phiphi}) gives
a shift at tree level in the gauge kinetic terms:
\be
\sum_{i = 1}^{3} \left( -\frac{1}{4} + \epsilon_i \right)
F^{\mu\nu}_i F_{i\, \mu\nu} ~,
\ee
with
\be
\epsilon_i \approx \frac{C_i}{M_s^2} \left|\langle \phi \rangle\right|^2 ~.
\label{dim6}
\ee
Thus, the bare gauge coupling $\bar{g}$ at the scale $M_s$ is shifted to
\be
g_i = \frac{\bar{g_i}}{\sqrt{1 - 4 \epsilon_i}} ~.
\label{barg}
\ee

Notice that the operator (\ref{phiphi}) is non-supersymmetric and
can be generated only below the supersymmetry breaking scale, $M_{\rm SSB}$.
If $M_{\rm SSB} < M_s$, then $C_i$ will be in fact suppressed by some power
of $M_{\rm SSB}/M_s$, depending on the dimensionality of the
supersymmetric operator that gives rise to the operator (\ref{phiphi})
below $M_{\rm SSB}$.

However, supersymmetric operators may also induce
significant shifts in the gauge couplings, provided
there are holomorphic combinations of superfields with scalar VEVs.
For example, in the MSSM, the operators
\be
\sum_{i = 1}^{3} \frac{C_i}{M_s^2}\int d^2\theta H_u H_d W_i W_i
\ee
gives
\be
\epsilon_i \approx \frac{C_i v^2\tan\beta}{2M_s^2(1+ \tan^2\beta)}~,
\label{MSSMepsilon}
\ee
where $v\approx 246$ GeV is the electroweak scale, and $\tan\beta > 1$.

The lower bound on $M_s$ is about 5 TeV, and is discussed in the Appendix.
Thus, for $C_i$ of order one,
the corrections to the gauge couplings from the Higgs VEVs
are typically below one percent.
However, larger corrections could be induced if the effective theory below
$M_s$ includes other scalars with VEVs.
For instance, if $\phi$ transforms as a singlet under
$SU(3)_C \times SU(2)_W \times U(1)_Y$
(or as an adjoint or other representations contained in the
product of two adjoints under the unified gauge group~\cite{Hill84,
Hall:1993kq}), then
there are operators similar with (\ref{phiphi}) but linear in
$\phi$.
This is the case in the 
NMSSM, where there is a gauge singlet chiral superfield,
$S$, whose
scalar component has a VEV $\langle S \rangle$ that induces a $\mu$-term.
Therefore, the supersymmetric operators
\be
\sum_{i = 1}^{3} \frac{C_i^\prime}{M_s}\int d^2\theta S W_i W_i
\ee
gives rise to a dielectric constant $\epsilon_i$ which is linear in
$\langle S \rangle/M_s$. Generically, $\langle S \rangle$ is of the order
of the $\mu$ parameter in the MSSM, so that corrections to the gauge couplings
of order 10\% are typical for the NMSSM.

In principle, the corrections from string dynamics could be non-universal
and large in certain models.
For example, if there is a unified gauge symmetry broken at the string scale,
the operators
\be
\sum_{i = 1}^{3}
{C_i^{\prime\prime}\over M_s} \langle S_i \rangle F_{i\,\mu\nu} F_i^{\mu\nu} ~,
\label{dim5op}
\ee
shift the gauge couplings by order one if the VEVs of the
singlets $\langle S_i \rangle$ are comparable to $M_s$. The
corrections need not respect the unified symmetry since it
is already broken ({\it e.g.}, these singlets can come from
an adjoint field whose VEV breaks the unified gauge
group~\cite{Hill84}). If that is the case, then
the contribution (\ref{dim5op}) should not be artificially
separated from the universal
one because they are comparable and both occur at the string scale.
It seems more reasonable to say that
the gauge couplings simply do not unify in this case.
Then, the apparent convergence of the three Standard Model
gauge couplings would just be an unfortunate coincidence.
It is appropriate to talk about gauge coupling unification
only if the  non-universal corrections from string dynamics
are small, suppressed by
either loops or small VEVs. This happens for example when the
unified gauge symmetry is broken by Wilson loops in the string
theory~\cite{Witten85}, as there is no adjoint field VEV to provide
the corrections (\ref{dim5op}).

It is interesting that the shifts in gauge couplings discussed here
occur in the presence of any VEV, and do not require necessarily
fundamental scalar fields. For example, a fermion condensate,
$\langle \bar{\psi}\psi \rangle$, has contributions to $\epsilon_i$
suppressed by $M_s^3$.

\subsection{Shifts in gauge couplings due to loops}

For any scalar $\phi$, with or without a VEV, the
operator (\ref{phiphi}) leads
to a shift in the gauge coupling due to loop effects.
If supersymmetry is broken at $M_s$, the operator (\ref{phiphi})
gives rise at one loop to a quadratically divergent contribution
to the coefficient of $F^{\mu\nu}_i F_{i\,\mu\nu}$ (Fig.~1).
\begin{figure}
\centering
\centerline{\epsfxsize=2.5in\epsfysize=2.5in\epsfbox{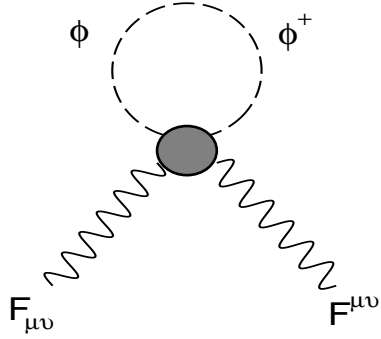}}
\vspace{-12mm}
\parbox{15cm}{\caption{\small
Loop contribution of the operator (\ref{phiphi}), represented by the
 blob, to the gauge kinetic term.
\label{loopdiag}}}
\end{figure}
This contribution has to be cut-off at $M_s$, resulting in a value
\be
\epsilon_i \approx \frac{n_s C_i}{(4\pi)^2} ~,
\label{eloop}
\ee
where $n_s$ is the number of complex degrees of freedom in $\phi$.

If the scalar $\phi$ has a VEV, both contributions (\ref{eloop})
and (\ref{dim6}) are present.
In the Standard Model, the higher-dimensional operators that
involve the Higgs doublet shift the gauge couplings as in
eq.~(\ref{barg}) with
\be
\epsilon_i \approx C_i \left(\frac{1}{8\pi^2} +
\frac{v^2}{2 M_s^2} \right) ~.
\label{Hdim6}
\ee
Values for $|\epsilon_i|$ of order $10^{-2}$
 are quite natural.
We emphasize that these corrections to the gauge couplings
are present in the Standard Model without the need of any new field
beyond the minimal content.

\section{Gauge Coupling Unification in Extra Dimensions}
\setcounter{equation}{0}

The gauge coupling constants in the $\overline{MS}$ scheme at the string
scale are related to the
measured coupling constants at the $Z$ pole by
\be
\overline{\alpha}_i^{-1}(M_s) =
\left[\alpha_i^{-1}(M_Z) - \frac{b_i}{2\pi} \ln \left(\frac{M_s}{M_Z}\right)
- \frac{\tilde{b}_i}{2\pi}
{\cal F}(\delta, R M_s) + \Delta_i^{\rm loops}\right]
\left(1 - 4 \epsilon_i \right)^{-1} ~,
\label{runcor}
\ee
where the $(1 - 4 \epsilon_i)^{-1}$ factor is due to the operators
suppressed by powers of $M_s$ (see Section 3),
${\cal F}(\delta, R \mu)$ is the one-loop contribution of the KK excitations
[see eq.~(\ref{run})].
The $\overline{\alpha}_i(\mu)$ represent the
coupling constants at the scale $\mu \le M_s$
in the absence of dimension-six or higher operators.

Gauge coupling unification is the condition
$\overline{\alpha}_1 = \overline{\alpha}_2 = \overline{\alpha}_3$ at $M_s$.
These two equations can be rewritten to leading order in $\epsilon_i$
and $\Delta_i^{\rm loops}$ as
\bear
\sum_{i,j,k = 1}^3 \varepsilon^{ijk} \left\{
\left(\tilde{b}_i - \tilde{b}_j \right) \alpha_{0_k}^{-1}
+ 4 \epsilon_k \left[\tilde{b}_i \alpha_{0_j}^{-1} -
\tilde{b}_j \alpha_{0_i}^{-1}         \right]
+  \left(\tilde{b}_i - \tilde{b}_j \right) \Delta_k^{\rm loops}   \right\}
= {\cal O}\left(\epsilon_i^2, \Delta_i^{\rm loops} \right)
\label{gendel} \\ [2mm]
{\cal F}(\delta, R M_s)  \approx  \frac{2\pi}{ \tilde{b}_3 - \tilde{b}_1 }
\left[ \alpha_{0_3}^{-1} - \alpha_{0_1}^{-1}
+ \Delta_3^{\rm loops} - \Delta_1^{\rm loops} +
\frac{4(\epsilon_1 - \epsilon_3) }{\tilde{b}_3 - \tilde{b}_1 }
\left(\tilde{b}_1 \alpha_{0_3}^{-1} - \tilde{b}_3 \alpha_{0_1}^{-1}
\right)\right]
\label{genf}
\eear
where $\alpha_{0_i}^{-1}$ represent the one-loop, inverse coupling
constants at $M_s$ in the absence of the KK excitations:
\be
\alpha_{0_i}^{-1} \equiv
\alpha_i^{-1}(M_Z) - \frac{b_i}{2\pi} \ln \left(\frac{M_s}{M_Z}\right) ~.
\ee
 Eq.~(\ref{gendel}) shows that if the sum over
$\left(\tilde{b}_i - \tilde{b}_j \right) \alpha_{0_k}^{-1}$ is much
smaller than the individual terms in this sum, then the corrections
(from two or more loops and from the string dynamics)
required by gauge coupling unification are indeed small.

The second unification condition, eq.~(\ref{genf}),
allows us to approximately determine the string scale
(assumed to be identical with the unification scale) as a function of
the compactification radius and the numbers of extra dimensions, for
any model in which the $b_i$ and $\tilde{b}_i$ coefficients
are consistent with eq.~(\ref{genf}) for small $\Delta_i^{\rm loops}$
and  $\epsilon_i$.
In the remainder of this section we analyze these unification
conditions in specific models.

\subsection{The Standard Model}

Consider the Standard Model in Minkowski spacetime
plus $\delta$ dimensions of radius $R \sim {\cal O}($TeV$^{-1})$,
with a string scale, $M_s$, above but not much larger than $1/R$.
In the four-dimensional Standard Model the one-loop
$\beta$-function coefficients, above $m_t$, are
\be
b_1 = \frac{41}{10}
\; , \; \; b_2 = - \frac{19}{6}  \; , \; \; b_3 = - 7 \; ~.
\ee
To avoid problems with chiral fermions in extra dimensions we assume
that the three generations of quarks and leptons are confined in the
three-dimensional flat space. Also, we assume that the Higgs doublet
cannot propagate in the compact dimensions. This situation can arise
due to orbifold compactifications in heterotic string theory,
or by duality, due to D-brane configurations in Type I string theory.
It can also arise in quantum field theory, if
there are domain walls in the $4+\delta$ dimensional theory.
In this case the KK excitations contribute at
each nondegenerate level to the $\beta$-function coefficients with
\be
\tilde{b}_1 = 0 \; , \; \; \tilde{b}_2 = - 7
\; , \; \; \tilde{b}_3 = - \frac{21}{2} ~.
\label{tildebsm}
\ee

We are now in a position to determine how large should be the
corrections from the string dynamics to the gauge couplings
in order to have gauge coupling unification.
Eq.(\ref{gendel}) gives
\be
\epsilon_3 - 1.5 \epsilon_2 + 0.6 \epsilon_1
+ 4.4\times 10^{-3} \left( \Delta_3^{\rm loops} - 1.5 \Delta_2^{\rm loops} +
0.5 \Delta_1^{\rm loops}\right) \approx 2.8 \times 10^{-2}
~,
\label{deltas}
\ee
for $M_s \lae {\cal O}(10^2)$ TeV.
This shows that indeed the corrections required by
gauge coupling unification are small, of order a few percent.
As explained in Section 3, values of order $10^{-2}$ for $|\epsilon_i|$,
as required by the unification condition (\ref{deltas}), are quite natural.

The other unification condition, eq.~(\ref{genf}), reads
\be
{\cal F}(\delta, R M_s)  \approx  25.4 \left[ 1
- 0.038 
\ln \left(\frac{M_s}{10 \; {\rm TeV}}\right)
+ 
0.024 \left( \Delta_1^{\rm loops} - \Delta_3^{\rm loops} \right)
+ 5.3 \left( \epsilon_1 - \epsilon_3 \right)
\right] ~.
\label{fvalue}
\ee
Using ${\cal F}(\delta, R M_s) \approx  25$, we compute $R M_s$ and
the number of KK modes from eq.~(\ref{fdel1}) for $\delta = 1$, and from
eq.~(\ref{calf1}) and Table 1 for $2\leq\delta\leq 6$. The results are
presented in Table 2.

\begin{table}[ht!]
\centering
\renewcommand{\arraystretch}{1.5}
\begin{tabular}{|c|r|r|c|c|} \hline
 $\;\delta\;$ &  $\;R M_s\;$ & $\;N_{\rm KK}\; $ & $n_{\rm max}$ &
$N_c N_{\rm KK}\alpha_i(M_s)/(4\pi)$ \\ \hline
\vbr\vbr
1 & $ 14.8  \;  $ &     $     28 \;$ & 15 & 0.12 \\
2 & $  4.2  \;  $  &    $     56 \;$ & 11 & 0.24 \\
3 & $  2.7  \;  $  &    $     80 \;$ & 6  & 0.35 \\
4 & $  2.2  \;  $  &    $     88 \;$ & 4  & 0.38 \\
5 & $  1.9  \;  $  &    $    130 \;$ & 3  & 0.56 \\
6 & $  1.8  \;  $  &    $    232 \;$ & 3  & 1.01 \\
\hline
\end{tabular}
\parbox{15.cm}{\caption{\small
The ratio of the string scale to the compactification scale, $ R M_s$,
and the corresponding number of KK modes, $N_{\rm KK}$, and KK levels,
$n_{\rm max}$, necessary for
gauge coupling unification in the Standard Model,
assuming ${\cal F}(\delta, R M_s) \approx 25$.
The loop expansion is reliable for $N_c N_{\rm KK}\alpha_i(M_s) \ll 4\pi$.
\label{nKK}}}
\end{table}

The phenomenological lower bound on the compactification scale,
$R^{-1} \gae 2.5$ TeV (see the Appendix), can be translated into a
lower bound on the string scale: $M_s \gae 40, \, 11, \, 7, \, 5$
TeV for $\delta = 1, \, 2, \, 3, \, \ge 4$, respectively.

{}From Table 2 one can see that the expansion parameter
$N_c N_{\rm KK}\alpha_i(M_s)/(4\pi)$
is not much smaller than one, so that perturbation theory is not
accurate, especially for more extra dimensions. Only for $\delta = 1$, this
expansion parameter may be sufficiently small, and it is reasonable to expect
corrections of order 10\% from two loops. Depending on the values of the
different parameters in eq.~(\ref{fvalue}), the expansion parameter may be
somewhat smaller than the values given in Table 2. For instance, for
$M_s \approx 40$ TeV, $\epsilon_2 = - \epsilon_1 = 10^{-2}$, and
small $\Delta_1^{\rm loops} - \Delta_2^{\rm loops}$, one gets
${\cal F}(\delta, R M_s) \approx 21.4$. For $\delta = 1$ this corresponds to
$N_{\rm KK} = 24$, and an expansion parameter of 0.10.

Nevertheless, we would like to determine more precisely
how large the higher order corrections are and how they
affect the unification. In the following we will estimate
the two-loop RG contributions of the KK states to the running
gauge couplings.

A problem of calculating higher loop corrections is that we
do not know a gauge invariant regularization method in higher
dimensional theories. A truncation of the KK modes at $M_s$
(just equivalent to the momentum cutoff in extra dimensions)
will not preserve gauge invariance, because the
gauge transformation~\cite{ddg}
\bear
\delta A_{\mu}^{a(n)} &=& \partial \theta^{a(n)} -{1\over 2}
f^{abc} \sum_m \left[ A_{\mu}^{b(n-m)} + A_{\mu}^{b(n+m)}\right]
\theta^{c(m)}, \\
\delta A_{5}^{a(n)} &=& -{n\over R} \theta^{a(n)} -{1\over 2}
f^{abc} \sum_m \left[ A_{5}^{b(n-m)} + A_{5}^{b(n+m)}\right]
\theta^{c(m)},
\eear
involves the sum of an infinite number of the KK states.
Without knowing a gauge invariant scheme, we will parameterize
our ignorance by some unknown parameter in the two-loop calculation.

The two-loop RG equations for the gauge couplings between $R^{-1}$
and $M_s$ are\footnote{This equation does not take into account
the difference between the zero mode and higher modes. However,
the difference is small and can be easily incorporated~\cite{ddg}.}
\be
\label{RGE}
{d\alpha_i^{-1}\over d (\ln \mu)} = - {\tilde{b}_i\over 2\pi}
N_{\rm KK}(\mu) - {\tilde{b^\prime}_i\over 2\pi} {\alpha_i\over 4\pi}
\xi N_{\rm KK}^2(\mu),
\ee
where $N_{\rm KK}(\mu)$ is the number of KK modes below the scale
$\mu$, $\tilde{b^\prime}_i$ are the two-loop RG coefficients for the field
content corresponding to one set
of KK states, $\tilde{b^\prime}_1=0$, $\tilde{b^\prime}_2=-36$,
$\tilde{b^\prime}_3 =-81$
\cite{vaug}, and $\xi$ parameterizes our ignorance of a suitable
gauge invariant regularization scheme. Naively one would expect
$0 < \xi <1$ because even if two independent KK state masses (momenta
in the extra dimensions) in a two-loop diagram are below $\mu$,
the other KK states in the two-loop diagram may have masses
greater than $\mu$, hence the corresponding diagram should not
be included (at least in a cutoff scheme). For simplicity, we
will work in the continuous limit since here we only concern about
the relative sizes of the one-loop and two-loop contributions.
In this limit, $N_{\rm KK}(\mu) \approx
(R\mu)^\delta \pi^{\delta/2}/\Gamma(1+\delta/2)$ is the volume
of a $\delta$-dimensional sphere of radius $R\mu$.

\begin{figure}[t]
\centering
\centerline{\hspace{3mm}\epsfxsize=2.5in\epsfysize=2.5in\epsfbox{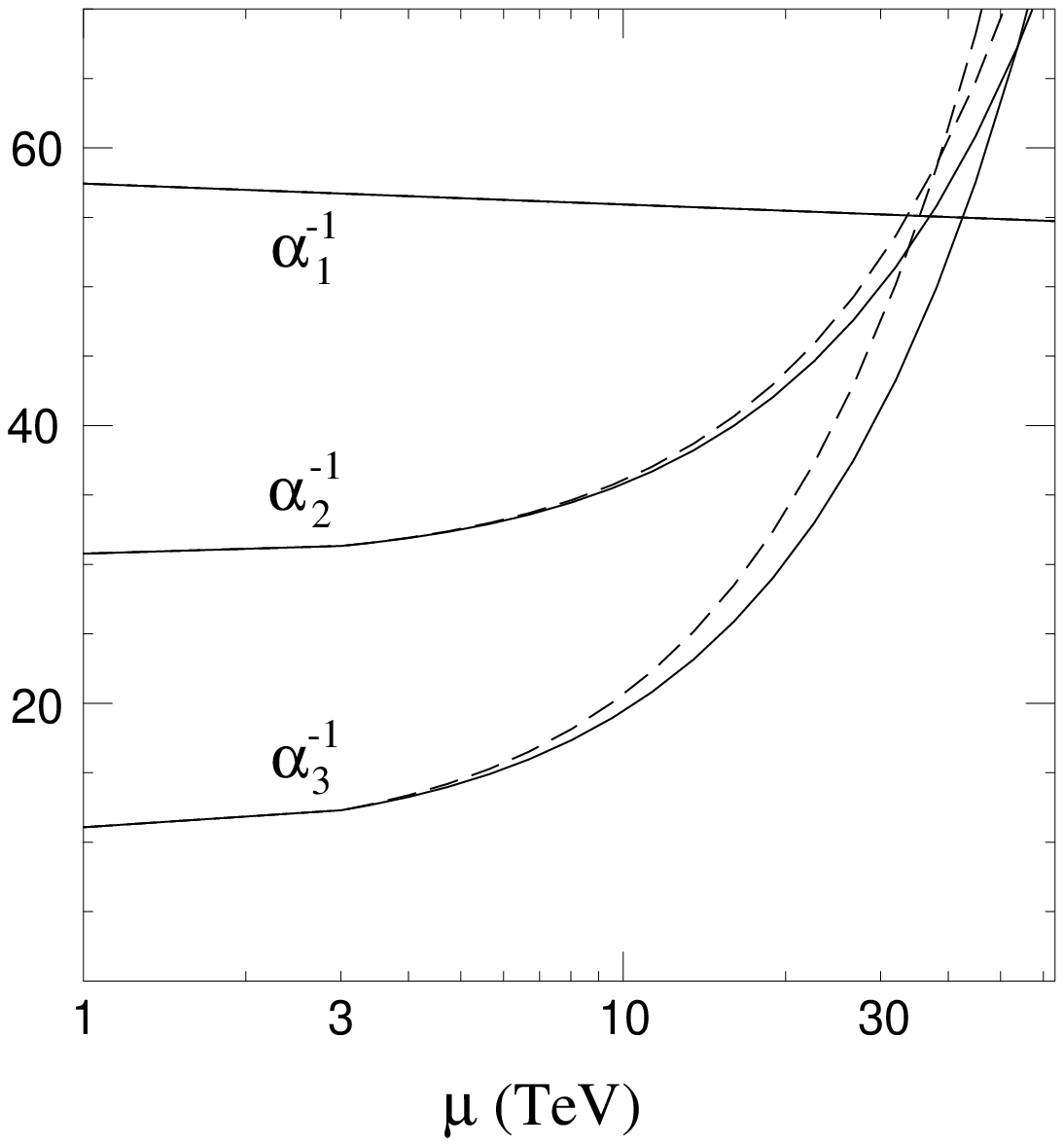}
\hspace{15mm}\epsfxsize=2.5in\epsfysize=2.5in\epsfbox{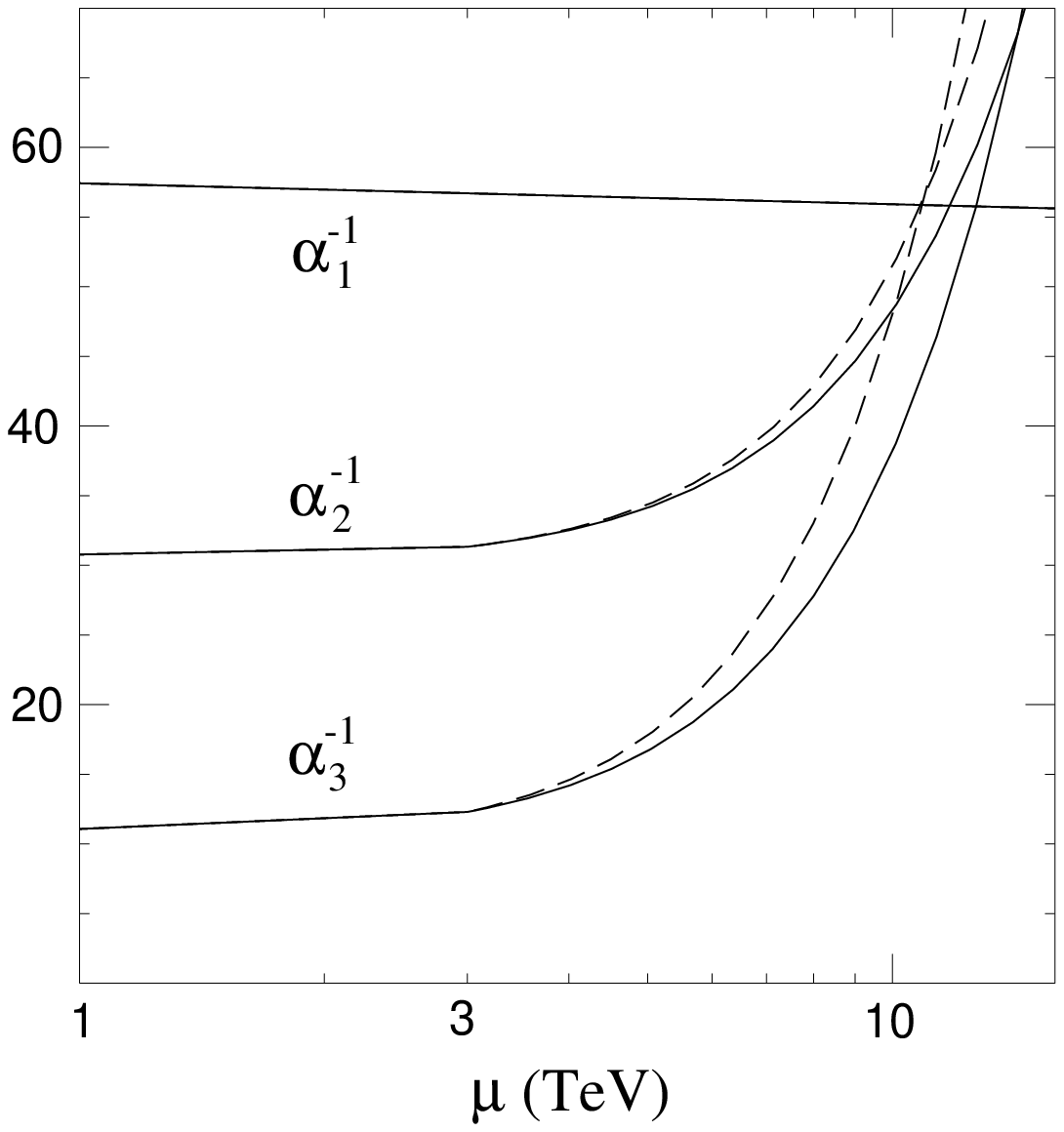}\hspace{9mm}}
\vspace{2mm}
\vspace{-1mm}
\parbox{15cm}{\hspace{33mm}{\bf a} \hspace{75mm} {\bf b}} \\ \vspace{3mm}
\parbox{14.5cm}{\small
\caption{\small Running of the SM gauge couplings for
a) $\delta=1$, and b) $\delta=2$ extra dimensions.
Solid curves are the 1-loop
results in the continuous limit. The dashed curves include the
two-loop contributions from the KK states, assuming $\xi=1$ in Eq.~(\ref{RGE}).
The KK excitations of the gauge bosons
do not affect $\alpha_1^{-1}$ directly because the gauge group is Abelian.
The compactification scale is assumed to be $1/R=3$ TeV.
\label{running1d}}}
\end{figure}

For a given compactification scale $R^{-1}$ we can solve (\ref{RGE})
numerically with the initial value $\alpha_i^{-1}(R^{-1})$ obtained
from the usual 4-dimensional running below $1/R$ (where one loop
is sufficient). The results are shown in Fig.~\ref{running1d}
 for $R^{-1}=3$ TeV, $\delta=1,\,2$, and $\xi=0$
(one-loop running) and $\xi=1$.

\begin{figure}[b!]
\centering
\vspace{-8mm}
\parbox{7.5cm}{\parbox{7.5cm}{ } \\ \vspace{11mm} \parbox{7.5cm}
{\hspace*{-6mm}\epsfxsize=2.9in\epsfysize=2.6in\epsfbox{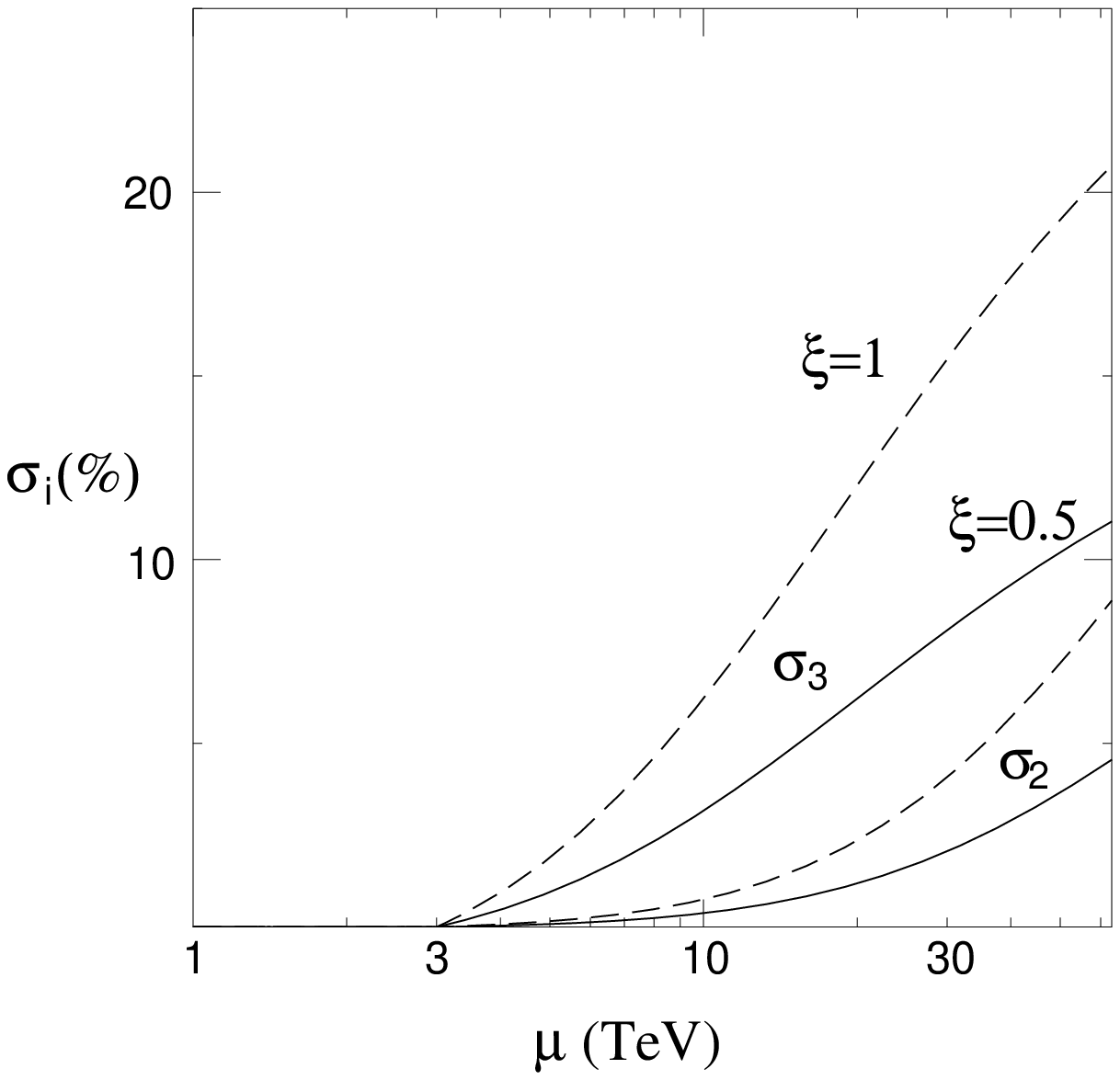}
 \hspace{10mm}}}
\parbox{7.5cm}{\hspace*{-8mm}\epsfxsize=3in\epsfysize=2.5in\epsfbox{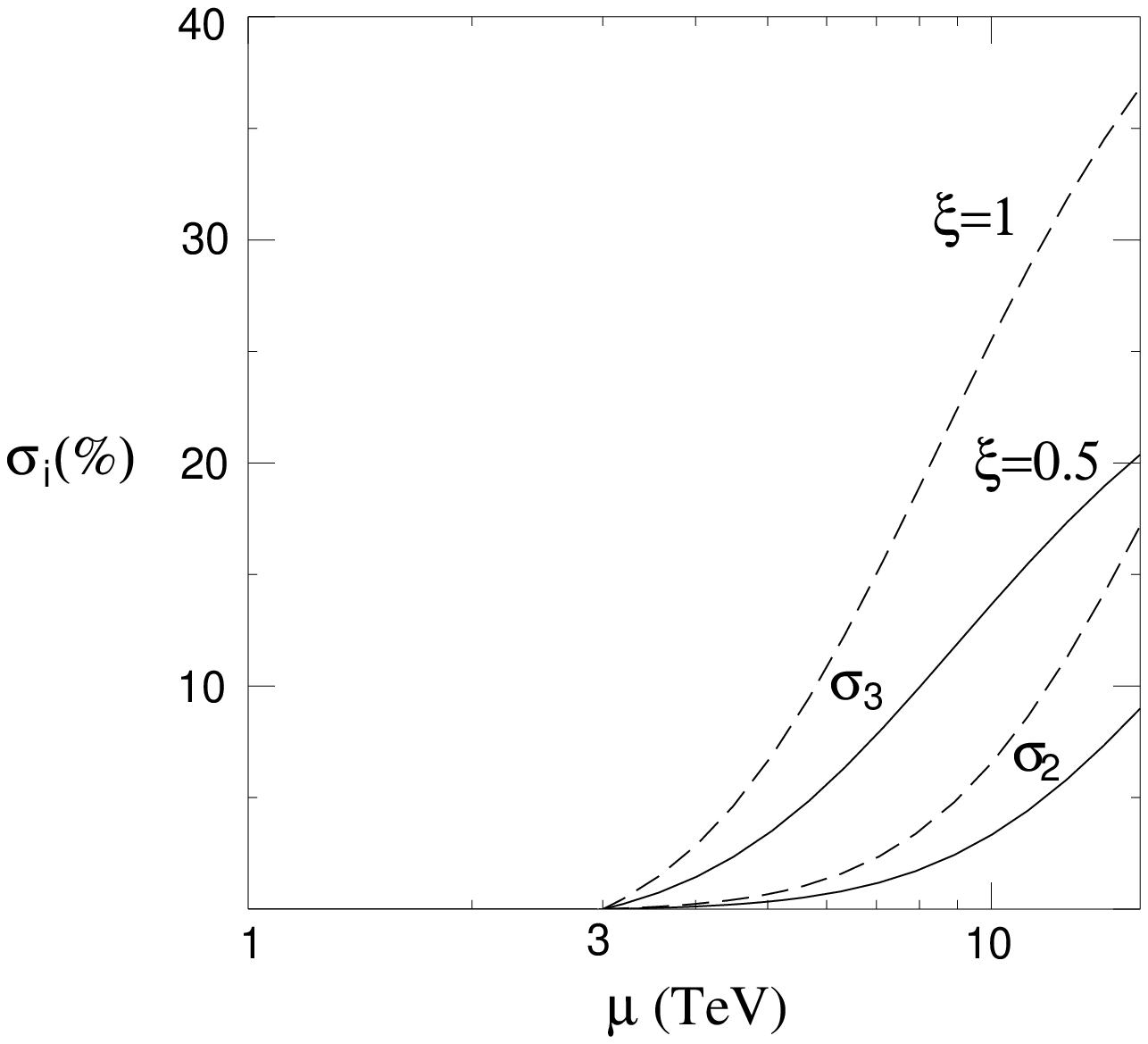}
 \hspace{15mm}} \\
\vspace{-5mm}
\parbox{15cm}{\hspace{36mm}{\bf a} \hspace{75mm} {\bf b}} \\ \vspace{3mm}
\parbox{14.5cm}{\small
\caption{\small The percentage corrections of $\alpha_2^{-1}$ and
$\alpha_3^{-1}$in the Standard Model due to the 2-loop  effects,
$\sigma_i = (\alpha_{i,\,{\rm
2-loop}}^{-1}/
\alpha_{i,\,{\rm 1-loop}}^{-1})(\mu)-1$, with $\xi=0.5$ (solid lines)
and $\xi=1$ (dashed lines), for: a) $\delta =1$, and b) $\delta =2$.
\label{corrections}}}
\end{figure}

We can see that the two-loop contributions improve the unification
for typical values of the unknown $\xi$ factor. Of course, this is sensible
only if the perturbation series converge and higher loop effects
are small. In Fig.~\ref{corrections} we plot the percentages of
corrections due to the two-loop effects.

For $\delta=1$ and $\xi=1$, the corrections
are $\sim 15\%$ for $\alpha_3^{-1}$ and $\sim 5\%$ for $\alpha_2^{-1}$.
The corrections are roughly proportional to $\delta$ and $\xi$. It
suggests that the perturbation series probably still holds for
small number of extra dimensions and may break down for more extra
dimensions. Note that the two-loop contributions reduce the gauge
couplings as well as the unification scale, so the expansion parameters
$\propto N_{\rm KK}\alpha_i$ are smaller than the one-loop estimate
(assuming that  $\xi > 0$).
Given possible corrections from string scale effects (discussed in
Section 3) and from higher loops, the Standard Model gauge couplings are
consistent with unification at the string scale without the need
of introducing new fields to change the $\beta$-functions
\cite{extramatter}, at least
for small number of extra dimensions where the perturbation series
can still be trusted.


\subsection{Higgs Compositeness from Extra Dimensions}

The condition of gauge coupling unification at one loop leads to the
conclusion that the gauge theory is in, or close to the non-perturbative
regime if there are more than one extra dimensions, as can be seen
by inspecting the values of the expansion parameter
$N_c N_{\rm KK}\alpha_i(M_s)/(4\pi)$ in Table 2.
The reason for that is that the higher-dimensional gauge coupling is
dimensionful such that the strength of the gauge interactions
increases rapidly above $R^{-1}$ \cite{AD}.

One possible effect of the non-perturbativity at the string scale is
the dynamical breaking of the chiral symmetry \cite{composite}.
If this is the case, then the fundamental Higgs doublet from the Standard
Model may be replaced with a composite Higgs, made up of the left-handed
top-bottom doublet and the right-handed component of a heavy vector-like
quark \cite{dhseesaw}.

The running of the gauge couplings in this case is almost the same as in
the Standard Model, assuming as in Section 4.1 that all the fermions are
confined on 3-branes.
The one-loop $\beta$-function coefficients due to a  nondegenerate set of
KK excitations is given by eq.~(\ref{tildebsm}).

The one-loop $\beta$-function coefficients due to zero modes are slightly
different than in the Standard Model. First, there is need for a vector-like
quark, $\chi$, with the same charges as the right-handed top.
Its mass $m_\chi$ is in the
TeV range, so that the logarithmic running due to this vector-like quark,
between $m_\chi$ and $M_s$, is negligible.
In addition to a composite Higgs doublet, other quark-antiquark bound states
may form, depending on the flavor structure of the four-quark operators
induced at the string scale \cite{composite}, and on the position of the
fermions in the extra dimensions \cite{AS}. These composite scalars
do not have KK modes, and contribute logarithmically to the gauge coupling
running between their masses and $R^{-1}$ (the compositeness scale is roughly
given by the compactification scale).
This ensures gauge coupling unification with a precision comparable
with the one in the Standard Model in extra dimensions.

Given that electroweak symmetry breaking occurs
only for more than one extra dimensions, where the dynamics
of the gluonic KK modes is strongly coupled,
it is not possible to predict reliably that gauge couplings unify in
this scenario. However,
it is interesting that, due to the presence of the extra dimensions,
gauge coupling unification is compatible with the formation of
a composite  Higgs doublet.

\subsection{Minimal Supersymmetric Standard Model}

We now turn to supersymmetric models in extra dimensions.
Supersymmetric operators induced at the string scale give
corrections to the gauge couplings of the order of the $SU(3)_C \times
SU(2)_W \times U(1)_Y$ invariant holomorphic
combinations of VEVs suppressed
by the appropriate power of $M_s$. As mentioned in Section 3,
in the MSSM these corrections are
likely to be below 1\%, while in the NMSSM
they may be as large as 10\%.

In the four-dimensional MSSM the one-loop
$\beta$-function coefficients above the superpartner masses are
given by
\be
b_1 = \frac{33}{5}
\; , \; \; b_2 = 1  \; , \; \; b_3 = - 3 \; ~.
\ee
Assuming as in Ref.~\cite{ddg} that only the vector multiplets and
the two Higgs supermultiplets propagate in the compact
dimensions, the  KK excitations give
\be
\tilde{b}_1 = \frac{3}{5} \; , \; \; \tilde{b}_2 = - 3
\; , \; \; \tilde{b}_3 = - 6 ~.
\ee

The unification conditions (\ref{gendel}) and (\ref{genf}) are given in
the MSSM by
\be
\epsilon_3 - 1.8 \epsilon_2 + 0.05 \epsilon_1
+ 5.0 \times 10^{-3} \left( \Delta_3^{\rm loops} - 1.8 \Delta_2^{\rm loops} +
0.83 \Delta_1^{\rm loops}\right) \approx - 1.4 \times 10^{-2}
~,
\label{mssmdeltas}
\ee
\be
{\cal F}(\delta, R M_s)  \approx  41.3 \left[ 1
- 0.035
\ln \left(\frac{M_s}{10 \; {\rm TeV}}\right)
+ 
0.023 \left( \Delta_1^{\rm loops} - \Delta_3^{\rm loops} \right)
+ 4.6 \left( \epsilon_1 - \epsilon_3 \right)
\right] ~.
\label{mssmfvalue}
\ee
for $M_s \lae {\cal O}(10^2)$ TeV.
The first equation here shows that the corrections required by
gauge coupling unification are again of order a few percent,
and slightly smaller than in the Standard Model.
In Section 3 we argued that the $\epsilon_i$ are below one percent in the
MSSM [see eq.~(\ref{MSSMepsilon})], and could be as large as 10\% in the
NMSSM.

Eq.~(\ref{mssmfvalue}) shows that ${\cal F}(\delta, R M_s)$,
and consequently the number of KK modes required by gauge coupling
unification, is larger than in the Standard Model. This is due to
the fact that the coefficients of the one-loop $\beta$-functions
are smaller in the MSSM. The number of KK modes, $N_{\rm KK}$,
and the loop expansion parameter $N_c N_{\rm KK}\alpha_i(M_s)/(4\pi)$
are listed in Table \ref{nKKMSSM}.
The expansion parameter appears to be too large (with the possible
exception of $\delta = 1$) to allow a reasonable convergence of the
perturbative series.

The KK excitations of the MSSM fields form complete ${\cal N} = 2$
supersymmetric multiplets \cite{ddg, zurab, Dienes:1998qh}. Since there is no
wave function renormalization beyond one-loop
in ${\cal N} = 2$ supersymmetric theories, it has been argued that
it is legitimate to study gauge coupling unification by keeping only
the one-loop running of the gauge couplings.
In fact there are higher-loop contributions to the gauge couplings
from Feynman diagrams involving at least one zero-mode \cite{zurab}.
These are subleading order in the large
$N_{\rm KK}$ limit, so that the one-loop contribution is indeed the largest
one. However, this does not improve the convergence of the perturbative
expansion, because all the terms in this series are of the same order
(albeit smaller than the one-loop term) for $\delta \gae 2$.
Therefore, we conclude based on the entries in last column of Table 3
that the corrections to the gauge couplings in the MSSM are large,
and if there is gauge coupling unification in the presence of more than one
compact dimension, then this happens in the non-perturbative regime.

\begin{table}[h!]
\centering
\renewcommand{\arraystretch}{1.5}
\begin{tabular}{|c|r|r|c|c|} \hline
 $\;\delta\;$ &  $\;R M_s\;$ & $\;N_{\rm KK}\; $ & $n_{\rm max}$ &
$N_c N_{\rm KK}\alpha_i(M_s)/(4\pi)$  \\ \hline
\vbr\vbr
1 & $ 22.5  \;  $ &     $     44 \;$ & 22 & 0.22 \\
2 & $  5.2  \;  $  &    $     88 \;$ & 14 & 0.44 \\
3 & $  3.1  \;  $  &    $    122 \;$ & 8  & 0.60 \\
4 & $  2.4  \;  $  &    $    136 \;$ & 5  & 0.68 \\
5 & $  2.1  \;  $  &    $    220 \;$ & 4  & 1.1 \\
6 & $  1.9  \;  $  &    $    232 \;$ & 3  & 1.2 \\
\hline
\end{tabular}
\parbox{15.4cm}{\caption{\small 
The ratio of the string scale to the compactification scale, $ R M_s$,
and the number of KK modes, $N_{\rm KK}$, and KK levels,
$n_{\rm max}$, required by
gauge coupling unification in the MSSM, for
${\cal F}(\delta, R M_s) \approx 40$.
The loop expansion works well for $N_c N_{\rm KK}\alpha_i(M_s) \ll 4\pi$.
\label{nKKMSSM}}}
\end{table}

Finally, we note that the gauge invariant operators
that are expected to be induced by the string dynamics have an impact on
four-dimensional supersymmetric models too.
If all these operators are generated with coefficients of order
one times the appropriate power of $M_s$, then any model that reduces
to the MSSM at low energy should include new gauge symmetries that
forbid the dangerous effects, such as proton decay, FCNC's, a large
$\mu$ term, and so on. The models of this type \cite{cdm}
do not accommodate gauge coupling unification in any obvious way.

\section{Conclusions}
\setcounter{equation}{0}

Our present analysis suggests the
following conjecture: theories in which there
are new dimensions at the $\sim$ TeV scale are
strongly coupled theories in the 
gauge couplings of $SU(3)_C \times SU(2)_W \times U(1)_Y$.
This has two important implications.  First, electroweak
symmetry breaking may be induce by this strong dynamics.
Much effort must now go into understanding strongly
coupled theories in extra-dimensions, and we defer this
to future studies.

The second implication, considered in the present paper, is that
potentially large uncertainties may affect gauge coupling unification.
We have shown (as an example, rather than a serious model proposal) that even
the minimal Standard Model is consistent with unification in the presence of 
compact dimensions and higher-dimensional operators that are expected 
to be induced by the string dynamics. 
Of course, unification is occurring at or near the string
scale where the precise physical meaning of unification is less clear.

The coefficients of the new higher dimensional operators cannot be
computed without a complete knowledge of and
computational capability within  the string theory.
Given these uncertainties we argue that the gauge coupling
unification is not a strong constraint on a large class of models. \\

{\it Acknowledgements:} We would like to thank Joe Lykken,
Konstantin Matchev, Stuart Raby, and Eric Weinberg for useful discussions.

\section*{Appendix: \ Lower Bounds on the Compactification and String Scales}
\renewcommand{\theequation}{A.\arabic{equation}}
\setcounter{equation}{0}

In this Appendix we discuss the lower bounds on the compactification
scale for the extra dimensions where Standard Model
 gauge fields propagate, and the string scale.

There are various experimental bounds on the compactification scale.
Direct searches for heavy gauge bosons put limits on the masses
of the KK states of the Standard Model gauge bosons.
It is useful to observe that the KK modes of the gluons have the same
couplings (up to an overall normalization) and properties as the
flavor-universal colorons \cite{colorons}.
By searching for new particles decaying to two-jets, CDF
gives a lower limit of 980 GeV on the flavor-universal
colorons~\cite{CDF}, and D0 put lower limits on additional Standard Model
$W$ and $Z$ bosons of 680 GeV and 615 GeV respectively~\cite{D0}.
Therefore, the compactification scale $R^{-1}$ has to be greater
than about 1 TeV.

There are also indirect limits from the electroweak
observables and higher dimensional operators
({\it e.g.,} four fermion operators) generated
from integrating out the heavy KK gauge bosons~\cite{hmass,NY,MP,Mar,ABQ}.
Many of them are related to the Fermi constant $G_F$,
since it is very precisely measured and is used as an
input parameter for determining other observables.
For example, in Ref.~\cite{NY}, it was claimed that from the
precision determination of $G_F$, one can exclude
the compactification scale below $\sim$ 1.6 TeV, 3.5 TeV,
5.7 TeV, and 7.8 TeV for 1, 2, 3, and 4 extra dimensions.
However, these indirect constraints are not as robust as
the constraints from direct searches, since
there may be some other unknown contributions to these
observables. As we have seen in the previous section,
the string scale $M_s$ is expected to be not far above the
compactification scale, due to the rapid convergence
of the gauge couplings (and also the rapid growing of the
't Hooft coupling $N_{\rm KK} g^2$) above the compactification
scale. In that case, we expect that there are
higher dimensional operators,
generated from the string scale physics, including those with
 the same form as the ones induced by the exchange of KK modes.
They are suppressed by a somewhat larger string scale mass, but they do not
have the small gauge coupling suppression as those coming from
integrating out KK gauge bosons. As a result, their sizes could
be comparable and they could cancel each other.

In addition, the corrections
from integrating out KK modes for some observables also depend
on how the four-dimensional theory is embedded into higher dimensions.
For example, the corrections to $G_F$ can be of opposite signs
depending on whether the Higgs propagates in the bulk or only on
the wall~\cite{MP}, and
whether the muon and the electron are localized at the same
point in the extra dimensions~\cite{AS}.
If different fermion species in a higher dimensional operator are
localized at different points in the compact dimensions, the KK
states will have different couplings to these fermions and the
sum of the contributions of all KK states may not add up.
This can give the right-sign correction
to the atomic parity violation~\cite{AS}. The strongest bound
comes from the leptonic width of the $Z$, $\Gamma(l^+ l^-)$, which
gives $R^{-1} \gae 2.5$ TeV for one extra dimension~\cite{MP}. However,
as we discussed above, this bound is still model dependent and may
be lowered by different arrangements of the Higgs
and leptons in the extra dimensions ({\it e.g.,} in the bulk or
on the wall, same point or different points, etc.)
and extra contributions to the electroweak parameters from the string
dynamics.

Now we turn to bounds on the string scale, $M_s$.
In addition to the effect of producing gauge dielectric constants,
discussed in Section 3,
the operators associated with a low string scale and the
extra dimensions at a TeV scale may have many new implications for
physics at lower energies. Some of these
operators have been analyzed in Ref.~\cite{HK}.

The fundamental theory incorporates quantum gravity, so that
it is  expected that string scale physics will generate higher dimensional
gauge invariant operators at $M_s$.
If all such higher dimensional operators suppressed by the
appropriate powers of the fundamental scale, $M_s$, are generated
with coefficients of order one, then the strong constraints from
proton decay, flavor changing processes, and CP violation will push
$M_s$ to a very high scale, reintroducing the hierarchy problem.
However, these problems are closely related to the pattern of
flavor symmetry breaking. They may have some higher dimensional
solutions~\cite{ddg,AS}, or can be avoided if there is a large
flavor symmetry~\cite{AD,HK}.
For flavor-conserving operators, the constraints coming from
compositeness searches on the four fermion operators require
that $M_s \gae 1$ TeV~\cite{ADD2,Cheung}. The strongest constraints
come from the precision measurements of the electroweak
sector~\cite{HK,BDN,BS}. The operators
\bear
\label{OBW}
{\cal O}_{BW} &=& {c_{BW}\over M_s^2} B^{\mu\nu} \left(
H^{\dagger}{\sigma^a \over 2} W^a_{\mu\nu} H\right),\\
\label{OH}
{\cal O}_{H} &=& {c_H \over M_s^2} \left( H^{\dagger}
{\cal D}^{\mu} H\right) \left( {\cal D}_{\mu} H^{\dagger} H
\right),
\eear
contribute to the $S$ and $T$ parameters
respectively~\cite{HISZ,HK}.
A global fit to the electroweak observables gives the
constraints~\cite{HK}
\bear
{M_s\over \sqrt{c_{BW}}} &>& 3.6 \;\rm{TeV},\\
{M_s\over \sqrt{c_{H}}} &>& 3.0 \;\rm{TeV} ~.
\eear
Note that these constraints apply at the electroweak scale while
the higher dimensional operators are generated at the string
scale. In running from the string scale down to the electroweak scale,
the coefficients of these operators are renormalized not only
by themselves but also form the operators (\ref{phiphi}).
Although they receive power law corrections between the
string scale and the compactification scale, we do not
expect more than a factor of 2 modifications due to the
smallness of $g_1$ and $g_2$, and closeness between the
two scales (for comparison, $\alpha_2$ changes by a factor
of about 2.) The one loop contributions of the operators
(\ref{phiphi}) to (\ref{OBW}) and hence to $S$ were
calculated in Ref.~\cite{HISZ}. They are roughly of the
order $0.1 \times (c_{WW},\, c_{BB})$, so the direct constraints
on the operators (\ref{phiphi}) from $S$ and $T$
are not very strong.
There are also various higher dimensional operators which give
non-universal contributions to individual observables. The
constraints depend sensitively on the sizes and signs of the
coefficients of these operators. Assuming that all coefficients
are $\pm 1$~\cite{BS}, $M_s \sim 5$ TeV is still allowed
for some choices of the signs.


\vfil

\begin{thebibliography}{99}
\frenchspacing

\bibitem{geogla} H.~Georgi and S.~L.~Glashow, Phys.~Rev.~Lett.~{\bf 32}
        (1974) 438.
\bibitem{mssm} J.~Ellis, S.~Kelley and D.V.~Nanopoulos,
Phys. Lett. {\bf B249} (1990) 441; \ Phys. Lett. {\bf B260} (1991) 131; \\
P.~Langacker and M.~Luo, Phys. Rev. {\bf D44} (1991) 817; \\
U.~Amaldi, W.~de Boer and H.~Furstenau, Phys. Lett. {\bf B260} (1991) 447; \\
C.~Giunti, C.W.~Kim and U.W.~Lee, Mod. Phys. Lett. {\bf A6} (1991) 1745; \\
M.~Carena, S.~Pokorski and C.E.~Wagner, Nucl. Phys. {\bf B406} (1993) 59,
hep-ph/9303202.
\bibitem{precision} J.~Bagger, K.~Matchev and D.~Pierce,
         Phys. Lett. {\bf B348}, 443 (1995), hep-ph/9501277; \\
        P.~Langacker and N.~Polonsky,
           Phys. Rev. {\bf D52}, 3081 (1995), hep-ph/9503214; \\
    D.M.~Pierce, J.A.~Bagger, K.~Matchev and R.~Zhang,
         Nucl. Phys. {\bf B491}, 3 (1997), hep-ph/9606211.
\bibitem{alphas} C. Caso, {\it et al}, Euro.~Phys.~Journal C3 (1998) 1.
\bibitem{threshold} L.~Hall,
         Nucl. Phys. {\bf B178}, 75 (1981).
\bibitem{Hill84} C.~T.~Hill, Phys.~Lett.~{\bf B135} 47(1984); \\
      Q. Shafi and C. Wetterich, Phys.~Rev.~Lett. {\bf 52} (1984) 875.
\bibitem{ddg} K.~R.~Dienes, E.~Dudas, and T.~Ghergheta,
        Phys.~Lett.~{\bf B436}, 55 (1998), hep-ph/9803466; \
        Nucl.~Phys.~{\bf B537}, 47 (1999), hep-ph/9806292.
\bibitem{Antoniadis:1990ew}
I.~Antoniadis,
Phys. Lett. {\bf B246} (1990) 377.
\bibitem{uni}
       D.~Ghilencea and G.~G.~Ross, Phys.~Lett.~{\bf B442}, 165 (1998),
      hep-ph/9809217; \\
        S.~A.~Abel and S.~F.~King, Phys.~Rev.~{\bf D59}, 095010 (1999),
      hep-ph/9809467; \\
       C.~D.~Carone, hep-ph/9902407; \\
	A.~Delgado and M~Quiros, hep-ph/9903400.
\bibitem{Perez-Lorenzana:1999qb}
A.~Perez-Lorenzana and R.N.~Mohapatra,
hep-ph/9904504.
\bibitem{AS} N.~Arkani-Hamed and M.~Schmaltz, SLAC-PUB-8082,
       hep-ph/9903417.
\bibitem{HK} L.~Hall and C.~Kolda, LBNL-43085, hep-ph/9904236.
\bibitem{hmass} T.~Rizzo and J.~Wells, hep-ph/9905234; \\
	A.~Strumia, hep-ph/9906266.
\bibitem{dine} M.~Dine and N.~Seiberg, Phys.~Lett.~{\bf B162} (1985) 299; \\
        M.~Dine, hep-ph/9905219.
\bibitem{lykken} J.~Lykken, Phys.~Rev.~{\bf D54}, 3693 (1996),
        hep-th/9603133.
\bibitem{largedim} N.~Arkani-Hamed, S.~Dimopoulos and G.~Dvali,
        Phys.~Lett.~{\bf B429}, 263 (1998), hep-ph/9803315; \\
I.~Antoniadis, N.~Arkani-Hamed, S.~Dimopoulos and G.~Dvali,
Phys. Lett. {\bf B436} (1998) 257.
hep-ph/9804398.
\bibitem{warp} L.~Randall and R.~Sundrum, MIT-CTP-2860, hep-ph/9905221.
\bibitem{composite}  B.~A.~Dobrescu, hep-ph/9812349 and hep-ph/9903407.
\bibitem{kacmoody}
For a review, see K.~R.~Dienes, Phys.~Reports {\bf 287} (1997) 447,
hep-th/9602045.
\bibitem{difbrane} G.~Shiu and S.-H.~H.~Tye, Phys.~Rev.~{\bf D58}, 106007
        (1998), hep-th/9805157.

\bibitem{ADD2} N.~Arkani-Hamed, S.~Dimopoulos and G.~Dvali,
        Phys.~Rev.~{\bf D59}, 086004 (1999), hep-ph/9807344.
\bibitem{moreuni}
	Z.~Kakushadze and S.-H.~H.~Tye, hep-th/9809147; \\
C.~P.~Bachas, hep-ph/9807415; \\
L. E. Ibanez, hep-ph/9905349; \\
I. Antoniadis, C.~Bachas, and E.~Dudas, hep-th/9906039.
\bibitem{Hall:1993kq}
L.J.~Hall and U.~Sarid, Phys. Rev. Lett. {\bf 70} (1993) 2673, hep-ph/9210240.
\bibitem{Witten85} E.~Witten, Nucl. Phys. {\bf B258} (1985) 75.

\bibitem{vaug}
M.E.~Machacek and M.T.~Vaughn, Nucl. Phys. {\bf B222}, 83 (1983).
\bibitem{extramatter} P.~H.~Frampton and A.~Rasin, IFP-769-UNC,
      hep-ph/9903479.
\bibitem{AD} N.~Arkani-Hamed and S.~Dimopoulos, SLAC-PUB-8008,
       hep-ph/9811353.
\bibitem{dhseesaw}
      B.~A.~Dobrescu and C.~T.~Hill, Phys.~Rev.~Lett.~{\bf 81} (1998) 2634,
      hep-ph/9712319; \\
R.~S.~Chivukula, B.~A.~Dobrescu, H.~Georgi, and C.~T.~Hill,
         Phys.~Rev.~{\bf D59} (1999) 075003, hep-ph/9809470.
\bibitem{zurab} Z. Kakushadze, HUTP-98-A073, hep-th/9811193 ;\\
Z. Kakushadze and T. R. Taylor, HUTP-99-A019, hep-th/9905137.
\bibitem{Dienes:1998qh}
K.~R.~Dienes, E.~Dudas and T.~Gherghetta,
hep-ph/9807522.
\bibitem{cdm} H.-C.~Cheng, B.~A.~Dobrescu, and K.~T.~Matchev,
         Nucl.~Phys.~{\bf B543} 47, (1999), hep-ph/9811316.

\bibitem{colorons} 
R.S.~Chivukula, A.G.~Cohen and E.H.~Simmons,
Phys. Lett. {\bf B380} (1996) 92,
hep-ph/9603311; \\
E.H.~Simmons,
Phys. Rev. {\bf D55} (1997) 1678,
hep-ph/9608269; \\
M.B.~Popovic and E.H.~Simmons,
Phys. Rev. {\bf D58} (1998) 095007,
hep-ph/9806287; \\
I.~Bertram and E.~H.~Simmons, Phys. Lett.~{\bf B443}
	(1998) 347, hep-ph/9809472.
\bibitem{CDF} F.~Abe, {\it et al.} (CDF Collaboration),
      Phys.~Rev.~{\bf D55}, 5263 (1997), hep-ex/9702004.
\bibitem{D0} B.~Abbott {\it et al.} (D0 Collaboration),
      {\it XVIII International Symposium on Lepton-Photon
       Interactions}, Hamburg, Germany, July 18-August 1, 1997,
      Fermilab-Conf-97/356-E.
\bibitem{NY} P.~Nath and M.~Yamaguchi, hep-ph/9902323 and
      hep-ph/9903298.
\bibitem{MP} M.~Masip and A.~Pomarol, CERN-TH/99-47,
       hep-ph/9902467.
\bibitem{Mar} W.~J.~Marciano, hep-ph/9902332 and
     hep-ph/9903451.
\bibitem{ABQ} I.~Antoniadis, K.~Benakli and M.~Quir\'{o}s,
     CERN-TH/99-128, hep-ph/9905311.
\bibitem{Cheung} K.~Cheung, UCD-HEP-99-10, hep-ph/9904510.
\bibitem{BDN} T.~Banks, M.~Dine, and A.~E.~Nelson,
       SCIPP-99/03, hep-th/9903019.
\bibitem{BS} R.~Barbieri and A.~Strumia, IFUP-TH/21-99,
        hep-ph/9905281.
\bibitem{HISZ} K.~Hagiwara, S.~Ishihara, R.~Szalapski and
      D.~Zeppenfeld, Phys.~Rev.~{\bf D48}, 2182 (1993); \\
K.~Hagiwara, T.~Hatsukano, S.~Ishihara and R.~Szalapski,
      Nucl.~Phys.~{\bf B496} (1997) 66, hep-ph/9612268.

\end{thebibliography}
\end{document}